\documentstyle [12pt]{article}

\def\K{{\cal K}}

\def\R{{\rm I\hspace{-.15em}R}}

\def\b{\begin{equation}}
\def\e{\end{equation}}
\def\bd{\begin{displaystyle}}
\def\ed{\end{displaystyle}}
\def\ba{\begin{array}}
\def\ea{\end{array}}

\def\bee{\begin{enumerate}}
\def\eee{\end{enumerate}}

\def\bes{\begin{eqnarray*}}
\def\ees{\end{eqnarray*}}
\def\be{\begin{eqnarray}}
\def\ee{\end{eqnarray}}

\def\le{\langle}
\def\re{\rangle}

\begin{document}

\title{Conformally  Invariant ``Massless'' Spin-2 Field\\ in de Sitter Universe }

\author{M. Dehghani$^{1,2}$, S. Rouhani$^3$, M.V. Takook$^{1}$\thanks{e-mail:
takook@razi.ac.ir} and M.R. Tanhayi$^4$}

\maketitle \centerline{\it $^1$Department of Physics, Razi
University, Kermanshah, Iran} \centerline{\it $^2$ Department of
Physics, Ilam University,
    Ilam, Iran}\centerline{\it $^3$Plasma Physics
Research Center, Islamic Azad University, Tehran, Iran}
   \centerline{\it $^4$4Department of Physics, Islamic Azad University, Central Branch, Tehran, Iran }

 \vspace{15pt}

\begin{abstract}

``Massless'' spin-2 field equation in de Sitter space, which is
invariant under the conformal transformation, has been obtained. The
frame work utilized is the symmetric rank-2 tensor field of the
conformal group. Our method is based on the group theoretical
approach and six-cone formalism, initially introduced by Dirac.
Dirac's six-cone is used to obtain conformally invariant equations
on de Sitter space. The solution of the physical sector of massless
spin-2 field (linear gravity) in de Sitter ambient space is written
as a product of a generalized polarization tensor and a massless
minimally coupled scalar field. Similar to the minimally coupled
scalar field, for quantization of this sector, the Krein space
quantization is utilized. We have calculated the physical part of
the linear graviton two-point function. This two-point function is
de Sitter invariant and free of pathological large distance
behavior.

\end{abstract}

\vspace{0.5cm} {\it Proposed PACS numbers}: 04.62.+v, 03.70+k,
11.10.Cd, 98.80.H \vspace{0.5cm}

\newpage
\section{Introduction}

Quantum field theory in de Sitter (dS) space-time has evolved as
an exceedingly important subject, studied by many authors in the
course of the past decade. This is due to the fact that most
recent astrophysical data indicate that our universe might
currently be in a dS phase. The importance of dS space has been
primarily ignited by the study of the inflationary model of the
universe and the quantum gravity. In Minkowski space-time, it is
well known that the massless fields propagate on the light-cone.
These fields are invariant under the conformal group $SO(2,4)$.
For spin $s\geq 1$ they are invariant under the gauge
transformation as well. In dS space, mass is not an invariant
parameter for the set of observable transformations under the dS
group $SO(1,4)$. Concept of light-cone propagation, however, does
exist and leads to the conformal invariance. ``Massless" is used
here in reference to the conformal invariance (propagation on the
dS light-cone). The term ``massive" fields is refereed to the
fields that in their minkowskian limit (zero curvature) reduce to
massive minkowskian fields alone. Indeed, the principal series of
UIRs admits a massive Poincaré group UIR in the limit H=0.

It has been shown that the ``massive'' and ``massless'' conformally
coupled scalar fields in dS space correspond to the principal and
complementary series representations of dS group, respectively
\cite{brmo}. The ``massive'' vector field in dS space has been
associated with the principal series, whereas ``massless'' field
corresponds to the lowest representation of the vector discrete
series representation in dS group \cite{gata}. The ``massive'' and
``massless'' spin-2 fields in dS space have been also associated
with the principal series and the lowest representation of the
rank-2 tensor discrete series of dS group, respectively
\cite{gasp,gagata,an}. The importance of the ``massless'' spin-2
field in the dS space is due to the fact that it plays the central
role in quantum gravity and quantum cosmology.

In the previous paper, the conformally invariant (CI) wave equations
for scalar and vector fields in dS space were obtained \cite{me}. We
are interested in the conformal invariance properties of
``massless'' spin-2 field in dS space, {\it i.e.} dS linear gravity.
In this paper, the ``massless'' spin-2 CI wave equation is obtained.
The frame work utilized here is the symmetric rank-2 tensor field.
Our method is based on group theoretical point of view and Dirac's
six-cone formalism and the conformal space is used to obtain the CI
equations. The concept of conformal space was used by Dirac
\cite{dir} to demonstrate the field equations for spinor and vector
fields in $1+3$ dimensional space-time in manifestly CI form. He
embedded Minkowski space as the hyper surface
$\eta_{ab}u^{a}u^{b}=0,\,\,(a,b=0,1,2,3,4,5),
\,\,\eta_{ab}=diag(1,-1,-1,-1,-1,1)$ in $R^6$. Then he extended the
fields by homogeneity requirements to the whole of the space of
homogeneous coordinates, namely $R^6$. This formalism developed by
Mack and Salam \cite{s6} and many others \cite{o7}. This approach to
conformal symmetry leads to the best path to exploit the physical
symmetry in contrast to approaches based on group theoretical
treatment of state vector spaces associated with the group. We use
this formalism to obtain CI wave equations in dS space. Many believe
that conformal invariance may be the key to the solution of the
problem of quantum gravity. The conformal invariance, and the
light-cone propagation, constitutes the basis for construction of
``massless'' field in dS space. For $s\geq 1$, the gauge invariance
provides an additional tool for analysis of this problem.

The organization of this paper and its brief outlook are as follows.
Section $2$ is devoted to a brief review of the dS ``massless''
spin-2 field equations in the ambient space. Section $3$ introduces
Dirac's manifestly covariant formalism of symmetric tensor fields on
the six-cone; in this section, conditions for the existence of CI
wave equations are given. Invariant subspace of fields are defined
by means of subsidiary conditions: transversality,
divergencelessness. Section 4 is devoted to the solutions of the
physical part of field equations. We show that this physical sector
can be written in terms of a polarization tensor and a massless
minimally coupled scalar field
$$ \K_{\alpha\beta}(x)= {\cal D}_{\alpha\beta}(x,\partial)\phi (x). $$
A Krein space quantization \cite{gareta, dere} becomes necessary to
circumvent the corresponding well known anomalies. In section $5$ we
calculate the two-point function ${\cal W}_{\alpha\beta
\alpha'\beta'}(x,x')$ in ambient space notations. It is particularly
shown that obtaining a covariant two-point function without infrared
divergence necessitates the use a Krein space field quantization.
Finally a brief conclusion and an outlook for further investigation
has been presented. We have supplied some useful identities and
mathematical details of calculations in the appendices. Finally in
appendix F, the two-point function is calculated in terms of the
intrinsic coordinates from it's ambient space counterpart.

\setcounter{equation}{0}
\section{de Sitter field equations}

The dS metric is a solution of the cosmological Einstein's equation
with positive constant $\Lambda$. It is conveniently described as a
hyperboloid embedded in a five-dimensional Minkowski space
\begin{equation}
X_H=\{x \in \R^5 ;x^2=\eta_{\alpha\beta} x^\alpha x^\beta
=-H^{-2}=-\frac{3}{\Lambda}\},\;\;\alpha,\beta=0,1,2,3,4,
\end{equation}
where $\eta_{\alpha\beta}=$ diag$(1,-1,-1,-1,-1)$. The dS metrics
reads
$$ds^2=\eta_{\alpha\beta}dx^{\alpha}dx^{\beta}=g_{\mu\nu}^{dS}dX^{\mu}dX^{\nu},\;\;\mu,\nu=0,1,2,3$$
where the $X^\mu$'s are  $4$ space-time intrinsic coordinates of the
dS hyperboloid. Any geometrical object in this space can be written
in terms of the four local coordinates $X^\mu$ (intrinsics) or in
terms of the five global coordinates $x^\alpha$ (ambient space).

The linearized gravitational wave equation in intrinsic coordinates
is \cite{hiko1,hiko2}:
$$ \frac{1}{2}(\Box_H
 h_{\mu\nu}-\nabla_{\mu}\nabla^{\rho}h_{\nu\rho}-\nabla_{\nu}\nabla^{\rho}h_{\mu\rho}+\nabla_{\mu}
\nabla_{\nu}h')$$
\b+\frac{1}{2}g^{dS}_{\mu\nu}(\nabla_{\lambda}\nabla_{\rho}h^{\lambda\rho}-\Box_{H}
h')+H^{2}(h_{\mu\nu}+\frac{1}{2}h' g^{dS}_{\mu\nu})=0\e where
 $\Box_{H} \equiv \nabla_\mu \nabla^\mu$ is the Laplace-Beltrami
operator on dS space and $h'=h_{\mu}^{\mu}$ is the trace of
$h_{\mu\nu}$ with respect to the background metric. Here,
$\nabla^\nu$ is the covariant derivative, and the indices are raised
and lowered by the background metric ($
g_{\mu\nu}=g^{dS}_{\mu\nu}+h_{\mu\nu}$). Not that the field equation
$(2.2)$ is invariant under the gauge transformation \b h_{\mu\nu}
\longrightarrow
h_{\mu\nu}^{gt}=h_{\mu\nu}+\nabla_{\mu}\zeta_{\nu}+\nabla_{\nu}\zeta_{\mu},\e
where $\zeta$ is an arbitrary vector field.

In the following the ambient space notations is used; in these
notations, the relationship with unitary irreducible representations
(UIR's) of dS group becomes straightforward because the Casimir
operators are easy to identify \cite{G8}. There are two Casimir
operators $$ Q^{(1)}_2=-\frac{1}{2}L^{\alpha\beta}L_{\alpha\beta}=
-\frac{1}{2}(M^{\alpha\beta}+S^{\alpha\beta})(M_{\alpha\beta}+S_{\alpha\beta}),$$
\b
 Q^{(2)}_2=-W_{\alpha}W^{\alpha}\,,\e
 where
$M_{\alpha\beta}=-i(x_{\alpha}\partial_{\beta}-x_{\beta}\partial_{\alpha})=
-i(x_{\alpha}\bar\partial_{\beta}-x_{\beta}\bar\partial_{\alpha})$
and
$W_{\alpha}=-\frac{1}{8}\epsilon_{\alpha\beta\gamma\sigma\eta}L^{\beta\gamma}L^{\sigma\eta}$,
in which the symbol $\epsilon_{\alpha\beta\gamma\sigma\eta}$ holds
for the usual antisymmetric tensor. The action of the spin generator
$S_{\alpha\beta}$ is defined by \cite{G8}
$$S_{\alpha\beta}\K_{\gamma\delta}=-i(\eta_{\alpha\gamma}\K_{\beta\delta}-\eta_{\beta\gamma} \K_{\alpha\delta} +
 \eta_{\alpha\delta}\K_{\beta\gamma}-\eta_{\beta\delta}
 \K_{\alpha\gamma}),$$
$\bar\partial_{\alpha}$ is the tangential (or transverse) derivative
on dS space,$$
\bar\partial_{\alpha}=\theta_{\alpha\beta}\partial^{\beta}=\partial_{\alpha}+H^2x_{\alpha}x\cdot\partial,\,\,\,
with\,\,\,x\cdot\bar\partial=0\,,$$ and $\theta_{\alpha\beta}$ is
the transverse projector
($\theta_{\alpha\beta}=\eta_{\alpha\beta}+H^2x_{\alpha}x_{\beta}\,$).

It has been shown that the field equation $(2.2)$ in the ambient
space reads as (\cite{G8} and appendix B) \b
(Q^{(1)}_2+6)\K(x)+D_2\partial_2 \cdot\K=0,\e where $\partial_2
\cdot \K_{\alpha}=\bar\partial\cdot \K_{\alpha}-H^2
x_{\alpha}{\K}'-\frac {1}{2}\bar\partial_{\alpha}{\K}'\,$ and the
operator $ D_{2} $ is the generalized gradient defined by \b D_2
K=H^{-2}{\cal S}(\bar
\partial - H^2 x)K, \e note that ${\cal S} $ is the symmetrizer
operator and $K$ is a vector field. It is clear that the field
equation ($2.5$) is invariant under the following gauge
transformation\b \K_{\alpha\beta} \longrightarrow
\K_{\alpha\beta}^{gt}=\K_{\alpha\beta}+D_{2}\Lambda_{g}.\e

The operator $Q^{(1)}_2$ commutes with the action of the group
generators and, as a consequence, it is constant in each UIR's. Thus
the eigenvalues of $Q^{(1)}_2$ can be used to classify the UIR's
{\it i.e.,}
\begin{equation}
(Q^{(1)}_2-\langle Q^{(1)}_2\rangle){\cal K}(x)=0.
\end{equation}
Following Dixmier \cite{ms5}, we get a classification scheme using a
pair $(p,q)$ of parameters involved in the following possible
spectral values of the Casimir operators :
\begin{equation}
Q^{(1)}_{p}=\left(-p(p+1)-(q+1)(q-2)\right)I_d ,\qquad\quad
Q^{(2)}_{p}=\left(-p(p+1)q(q-1)\right)I_d\,.
\end{equation}
Three types of scalar, tensorial or spinorial UIR's are
distinguished for $SO(1,4)$ according to the range of values of the
parameters $q$ and $p$ \cite{ms5,ms6}, namely: the principal, the
complementary and the discrete series. The flat limit indicates that
for the principal and the complementary series value of $p$ bears
meaning of spin. For the discrete series case, the only
representation which has a physically meaningful Minkowskian
counterpart is $p=q$ case. Mathematical details of the group
contraction and the physical principles underlying the relationship
between dS and Poincar\'e groups can be found in Ref.s \cite{ms7}
and \cite{ms8} respectively. The spin-$2$ tensor representations
relevant to the present work are as follows:
\begin{itemize}
\item[i)] The UIR's $U^{2,\nu}$ in the principal series where
$p=s=2$ and $q=\frac{1}{2 }+i\nu$ correspond to the Casimir spectral
values:
\begin{equation}
\langle Q_2^{\nu}\rangle=\nu^2-\frac{15}{4},\;\;\nu \in \R,
\end{equation}
note that $U^{2,\nu}$ and $U^{2,-\nu}$ are equivalent. \item[ ii)]
The UIR's $V^{2,q}$ in the complementary series where $p=s=2$ and
$q-q^2=\mu,$ correspond to
\begin{equation}
\langle  Q_2^{\mu}\rangle=q-q^2-4\equiv
\mu-4,\;\;\;0<\mu<\frac{1}{4}\,.
\end{equation}
\item[iii)] The UIR's $\Pi^{\pm}_{2,q}$ in the discrete series
where $p=s=2$ correspond to
\begin{equation}
\langle Q_2^{(1)}\rangle=-4, \;\;q=1\; (\Pi^{\pm}_{2,1});
\;\;\langle Q_2^{(2)}\rangle=-6,\;\; q=2 \;(\Pi^{\pm}_{2,2}).
\end{equation}
\end{itemize}
The ``massless'' spin-2 field in dS space corresponds to the
$\Pi^{\pm}_{2,2}$ and $\Pi^{\pm}_{2,1}$ cases in which the sign
$\pm$, stands for the helicity. In these cases, the two
representations $\Pi^{\pm}_{2,2}$, in the discrete series with
$p=q=2$, have a Minkowskian interpretation. It should be noted that
$ p $ and $ q $ do not bear the meaning of mass and spin. For
discrete series in the limit $ H \rightarrow 0, $ $ p=q=s $ are
indeed none other than spin. The representation $\Pi^+_{2,2}$ has a
unique extension to a direct sum of two UIR's $ {\cal C}(3;2,0)$ and
${\cal C}(-3;2,0)$ of the conformal group $SO(2,4)$ with positive
and negative energies respectively \cite{ms7,ms9}. The latter
restricts to the massless tensor Poincar\'e UIR's ${\cal P}^>(0, 2)$
and ${\cal P}^<(0,2)$ with positive and negative energies
respectively. The following diagrams illustrate these connections
\begin{equation}
\left.
\begin{array}{ccccccc}
&& {\cal C}(3,2,0)& &{\cal C}(3,2,0)&\hookleftarrow &{\cal P}^{>}(0,2)\\
\Pi^+_{2,2} &\hookrightarrow &\oplus&\stackrel{H=0}{\longrightarrow} & \oplus  & &\oplus\\
&& {\cal C}(-3,2,0)& & {\cal C}(-3,2,0) &\hookleftarrow &{\cal
P}^{<}(0,2),\\
\end{array}
\right.
\end{equation}

\begin{equation}
\left.
\begin{array}{ccccccc}
&& {\cal C}(3,0,2)& &{\cal C}(3,0,2)&\hookleftarrow &{\cal P}^{>}(0,-2)\\
\Pi^-_{2,2} &\hookrightarrow &\oplus&\stackrel{H=0}{\longrightarrow}&\oplus &&\oplus\\
&& {\cal C}(-3,0,2)&& {\cal C}(-3,0,2)&\hookleftarrow &{\cal P}^{<}(0,-2),\\
\end{array}
\right.
\end{equation} where the arrows $\hookrightarrow $ designate unique
extension; $ {\cal P}^{ \stackrel{>} {<}}(0,2)$ (resp. $ {\cal
P}^{\stackrel{>}{<}}(0,-2)$) are the massless Poincar\'e UIR's with
positive and negative energies and positive (resp. negative)
helicity. It is important to note that the representations
$\Pi^{\pm}_{2,1}$ do not have corresponding flat limit.

\setcounter{equation}{0}
\section{Dirac's six-cone, conformally invariant equations}

In the Minkowski space, for every massless representation of
Poincar\'e group there exists only one corresponding representation
in the conformal group \cite{ms9,co2}. In the dS space, as
mentioned, for the ``massless'' tensor field, only two
representations in the discrete series $(\Pi^{\pm}_{2,2})$ have a
Minkowskian interpretation. The signs $\pm$ correspond to two types
of helicity for the ``massless'' tensor field. In this section, the
conformal invariance of ``massless'' tensor field in dS space is
studied. CI wave equations in dS space are best obtained by first
establishing the wave equations in Dirac's null-cone in $\R^6$, and
then  followed by the projection of these equations to the dS space.

Dirac's six-cone (or Dirac's projection cone) is defined  by $u^{2}
\equiv u^{2}_{0}-\vec u^{2}+u^{2}_{5}=0$, where $ \;u_{a} \in
\R^{6},$ and  $ \vec u \equiv(u_{1},u_{2},u_{3},u_{4})$. Reduction
to four dimensional (physical space-time) is achieved by projection,
that is by fixing the degrees of homogeneity of all fields. Wave
equations, subsidiary conditions, etc., must be expressed in terms
of operators that are defined intrinsically on the cone. These are
well-defined operators that map tensor fields on tensor fields with
the same rank on cone $u^2=0$ \cite{me,fr}. It is important to note
that on the cone $u^2=0$, the second order Casimir operator of
conformal group, ${\cal Q}_2 $, is not a suitable operator to obtain
CI wave equations. For example, for a symmetric tensor field of
rank-2, we have \cite{ms9,fr,co5}: \b {\cal
Q}_2\Psi^{cd}=\frac{1}{2} L_{ab}L^{ab}\Psi^{cd}=\left(-u^2
\partial^2+\hat{N_5}(\hat{N_5}+4)+8\right)\Psi^{cd},\e where $\hat{N_5}$ is
the conformal-degree operator defined by:\b \hat{N_5}\equiv
u^{a}\partial_{a}.\e On the cone this operator reduces to a
constant, {\it i.e.} $\hat{N}_5(\hat{N}_5+4)+8$. It is clear that
this operator cannot lead to the wave equations on the cone. The
well-defined operators exist only in exceptional cases. For tensor
fields of degree $ -1,0,1,...$, the intrinsic wave operators are $
\partial^2, (\partial^2)^2, (\partial^2)^3,...$ respectively
\cite{fr}. Thus the following CI system of equations, on the cone,
has been used \cite{me}: \b \left\{ \ba{rcl}
(\partial_a\partial^a)^n \Psi&=&0,\\
\hat{N_5}\Psi&=&(n-2)\Psi.\ea\right. \e where $\Psi$ is a tensor
field of a definite rank and of a definite symmetry.

Other CI conditions can be added to the above system in order to
restrict the space of the solutions. The following conditions are
introduced to achieve the above goal:
\begin{enumerate}

\item[a)]{transversality} $$u_a\Psi^{ab...}=0 ,$$

\item[b)]{divergencelessness} $$Grad_a\Psi^{ab...}=0 ,$$

\item[c)]{tracelessness} $$\Psi_{ab...}^a=0 .$$
\end{enumerate}

The operator $Grad_a$ unlike $\partial_{a}$ is intrinsic on the
cone, and is defined by \cite{fr}:\b Grad_a\equiv
u_a\partial_b\partial^b-(2\hat{N_5}+4)\partial_a \,.\e

In order to project the coordinates on the cone $u^2=0$, to the
$1+4$ dS space we chose the following relation: \b \left\{ \ba{rcl}
x^{\alpha}&=&(Hu^5)^{-1}u^\alpha,\\
x^5&=&Hu^5.\ea\right.\e Note that $x^5$ becomes superfluous when we
deal with the projective cone. It is easy to show that various
intrinsic operators introduced previously now read as:
\begin{enumerate}

\item{ the conformal-degree operator $(\hat{N_5})$} \b \hat{N_5}=
 x_5\frac{\partial}{\partial x_5},\e

\item{the conformal gradient $(Grad_{\alpha})$}  \b Grad_{\alpha}=
-x_{5}^{-1}
\{H^2x_{\alpha}[Q_{0}-\hat{N_5}(\hat{N_5}-1)]+2\bar{\partial}_{\alpha}(\hat{N_5}+1)\},\e

\item{and the powers of d'Alembertian $(\partial_a\partial^a)^n$},
which acts intrinsically on field of conformal degree $(n-2)$, \b
(\partial_{a}
\partial^{a})^n=-H^{2n}x_{5}^{-2n} \prod_{j=1}^{n}[Q_{0}+(j+1)(j-2)]\,. \e
\end{enumerate}
Considering the conformal invariance in the dS space, we classify
the $21$ degrees of freedom of the symmetric tensor field
$\Psi^{ab}$ on the cone by (in the following for the brevity we take
H=1)  \b {\cal K}_{\alpha\beta}=\Psi_{\alpha\beta}+{\cal
S}x_{\alpha}\Psi_{\beta}\cdot x+x_{\alpha}x_{\beta}x\cdot \Psi \cdot
x,\e $$ K_{\alpha}=x\cdot \Psi_{\alpha}+x_{\alpha}x \cdot \Psi \cdot
x,$$ $$ \phi= x\cdot\Psi \cdot x,$$ where $ \K_{\alpha\beta}$ and
$K_{\alpha}$ are tensor and vector fields on dS space respectively
$(x^\alpha \K_{\alpha\beta}=x^\beta \K_{\alpha\beta}=0 = x^\alpha
K_\alpha)$. The fields $
\Psi_{55},\,\,x.\Psi_{5},\,\,x_{\alpha}x.\Psi_{5}+\Psi_{\alpha 5}$
are auxiliary fields which are unnecessary to demonstrate on the dS
space.

In the following CI wave equation for the symmetric rank-2 tensor
field is considered. We have shown \cite{me}, for scalar and vector
fields, the simplest CI system of equations is obtained through
setting $n=1$ in $(3.3)$, {\it i.e.} the field with conformal-degree
$-1$, resulting equations are the UIR's of $SO(1,4)$. For a
symmetric tensor field of rank-2, the CI system $(3.3)$ with $n=1$
leads to (appendix B): \b (Q_0-2)\K_{\alpha\beta}+\frac{2}{3}{\cal
S}(\bar\partial_{\beta}+2x_{\beta})\bar\partial\cdot
\K_{\alpha}-\frac{1}{3}\theta_{\alpha\beta}\bar\partial\cdot\bar\partial\cdot\K=0\,.\e
It is clear that $(3.10)$ does not correspond to any UIR's of the dS
group. The intrinsic counterpart of $(3.10)$ becomes (appendix B) :
\b (\Box +4)h_{\mu\nu}-\frac{2}{3} {\cal S}\,\, \nabla_{\mu}\nabla
\cdot h_{\nu}+\frac{1}{3}g^{dS}_{\mu\nu}\nabla\cdot\nabla\cdot h=0,
\e in which the intrinsic field $h_{\mu\nu}$ is locally determined
by the transverse tensor field $\K_{\alpha\beta}$ through
$$h_{\mu\nu}(X)=\frac{\partial x^{\alpha}}{\partial
X^{\mu}}\frac{\partial x^{\beta}}{\partial X^{\nu}}
\K_{\alpha\beta}(x(X)).$$ Taking the flat limit $(H\rightarrow 0)$
of (3.11) we will gain the second order CI massless spin-$2$ wave
equation in four dimensional Minkowski space, which was found by
Barut and Xu \cite{Ba}. They have found the conformally covariant
massless spin-2 field equation by varying the coefficients of
various terms in the standard equation.

 In order to obtain CI wave equation for massless spin-2 field
which is physical state of dS space, let us set $n=2$ in (3.3), then
we have \b \left\{ \ba{rcl}
(\partial_a\partial^a)^2\Psi&=&0,\\
\hat{N_5}\Psi&=&0.\ea\right.\e \\
The following CI conditions can be added to the above system to
restrict the space of solutions:
\begin{enumerate}
\item[a)]{ transversality} $ u_a\Psi^{ab}=0 ,$ that results in \b
x^{5}(\Psi_{5b}+x\cdot \Psi_{b})=0,\e \item[b)]{ divergencelessness}
\b Grad_a\Psi^{ab}=0 .\e
\end{enumerate}

It is easy to show that (appendix C): \b \bar
\partial \cdot {\cal K}_{\alpha}=4(x \cdot
\Psi_{\alpha}+x_{\alpha}x \cdot \Psi \cdot x)\equiv 4K_{\alpha}. \e
then we get the following relation for the vector field
$\bar\partial \cdot {\cal K}_{\alpha}$ : \b Q_{1}\bar
\partial . {\cal K}_{\alpha}+\frac{2}{3}D_{1\alpha}\bar \partial
.\bar\partial .{\cal K} +\frac{1}{6}Q_{1}D_{1\alpha}\bar \partial
.\bar\partial .{\cal K}=0, \e where $D_1=-\bar \partial.$
 This CI equation is similar to the gauge-fixed wave equation for the vector field
$\bar\partial \cdot {\cal K}_{\alpha}$ \cite{me,in}. We are now in a
position to write CI system for dS field $\K_{\alpha\beta}$. In
order to accomplish this, we first determine the CI equation that
corresponds to UIR's of dS group (appendix D): \b
(Q_{2}+4)[(Q_{2}+6) {\cal K}_{\alpha\beta}+D_{2\alpha}\partial_2 .
{\cal K}_{\beta}]+ \frac {1}{3}D_{2\alpha}D_{1\beta} \bar\partial .
\bar
\partial . {\cal K}- \frac{1}{3}
\theta_{\alpha\beta}(Q_{0}+6)\bar\partial . \bar \partial . {\cal
K}=0. \e Finally CI system that obtained from $(3.12)$ with respect
to $(3.9)$ defined by

$$ (Q_{2}+4)[(Q_{2}+6) {\cal
K}_{\alpha\beta}+D_{2\alpha}\partial_2 . {\cal K}_{\beta}]+ \frac
{1}{3}D_{2\alpha}D_{1\beta} \bar\partial . \bar \partial . {\cal K}-
\frac{1}{3} \theta_{\alpha\beta}(Q_{0}+6)\bar\partial . \bar
\partial . {\cal K}=0,$$ $$Q_{1}\bar
\partial . {\cal K}_{\alpha}+\frac{2}{3}D_{1\alpha}\bar \partial
.\bar\partial .{\cal K} +\frac{1}{6}Q_{1}D_{1\alpha}\bar \partial
.\bar\partial .{\cal K}=0,$$ \b \K'=0\,.\e

It is important to note that by imposing the following conditions on
the tensor field ${\cal K}_{\alpha\beta}$,(which are necessary for
the UIR's of dS group)
$${\cal K}'=0=\bar\partial.{\cal K},$$ the CI system $(3.18)$ becomes $$ (Q_{2}+4)(Q_{2}+6) {\cal K}_{\alpha\beta}=0.$$
It is clear that this conformally invariant field corresponds to the
two representations of discrete series, namely $\Pi^{\pm}_{2,1}$ and
$\Pi^{\pm}_{2,2}$ (Eq. $(2.12))$, in other words it is the physical
representation of dS group. At this point it is clear that the
parameter $ p $ does have a physical significance. It is indeed
spin. In the following, we only consider the tensor field that
corresponds to the representations of discrete series
$\Pi^{\pm}_{2,2}$ which has the Minkowskian limit {\it i.e.} \b
(Q_{2}+6) {\cal K}_{\alpha\beta}=0,\,\,\bar\partial.{\cal K}=0={\cal
K}' .\e

\setcounter{equation}{0}
\section{de Sitter field solutions}

 The general solution of  Eq.$(3.19)$ can be written in the following form \cite {gagata,garota2007}
\b \K=\theta\phi_1+ {\cal S}\bar Z_1K+D_2K_g,\e where $ Z_{1} $ is a
constant 5-dimensional vector, $ \phi_1 $ is a scalar field, $ K $
and $ K_g $ are two vector fields. By using divergenceless
 and transversality conditions, we obtain $ {\cal K}'=0, $ which
results in \b 2\phi_1+Z_1.K+\bar{\partial}.K_g=0. \e Conditions $
x.K =0=x.K_g$ are used to obtain the above equation. By substituting
$ \K_{\alpha\beta} $ in $(3.19)$ we have \cite{gagata} \b \left\{
\begin{array}{ll}
          (Q_0+6)\phi_1=-4Z_1.K,\;\;\;\;\;\;(I)\\
          \\(Q_1+2)K=0, \;\;\;\;\;\;\;\;\;\;\;\;\;\;\;\;\;\;(II)\\
          \\(Q_1+6)K_g=2(x.Z_1)K.\;\;\;\;\;(III)
\end{array}\right.\e Using conditions $ x.K = 0=\bar{\partial}.K  $,
Eq.$(4.3-II)$ reduces  to $ Q_0 K=0$. From this reduced form and
Eq.$(4.3-I)$, we can write \b \phi_1=-\frac{2}{3}Z_1.K
\;\;\;,\;\;\;Q_0 \phi_1=0, \e and from Eq.$(4.2)$, we have \b
\bar{\partial}.K_g =\frac{1}{3}Z_1.K.\e

We choose the following form for the vector field $ K $ (the
solution of $(4.3-II)$) \cite {gata,gagarota} \b K=\bar Z_2
\phi_2+D_1 \phi_3,\e where $ Z_2 $ is another 5-dimensional constant
vector, $ \phi _2 $ and $ \phi_3 $ are two scalar fields.
Substituting $K$ into $(4.3-II)$ results in \b Q_0 \phi_2 =0. \e It
is clear that $ \phi_2 $ is a massless minimally coupled scalar
field. Using the divergenceless condition, $\phi_3$ can be written
in terms of $\phi_2$ \b \phi_3 =-\frac{1}{2} [Z_2.\bar
\partial \phi_2+2 x.Z_2\phi_2].\e So we can write \b K=\bar
Z_{2}\phi_2-\frac{1}{2}D_{1}[Z_2.\bar
\partial \phi_2+2 x.Z_2\phi_2],\e
 and
\b \phi_1 =-\frac{2}{3}Z_1.\left(\bar Z_{2}\phi_2-\frac{1}{2}D_{1}[
Z_2.\bar
\partial \phi_2+2 x.Z_2\phi_2]\right). \e
According to the following identity ( appendix E)\b
(Q_1+6)^{-1}(x.Z_1)K=\frac{1}{6}\left[(x.Z_1)K+\frac{1}{9}D_1(Z_1.K)\right],
\e the Eq.$(4.3-III)$ leads to \b
K_g=\frac{1}{3}\left[(x.Z_1)K+\frac{1}{9}D_1(Z_1.K)\right],\e where
$ x.K_g=0$ and $\bar{\partial}.K_g =\frac{1}{3}Z_1.K.$

Using the Eq.s $(4.9)$, $(4.10)$ and $(4.12)$, we can rewrite
$\K_{\alpha\beta}$ as the following form \b \K_{\alpha
\beta}(x)={\cal D}_{\alpha \beta}(x,\partial,Z_1,Z_2)\phi_2,\e where
${\cal D}$ is the projector tensor $$ {\cal
D}(x,\partial,Z_1,Z_2)=\left(-\frac{2}{3}\theta Z_1.+{\cal S}\bar
Z_1+\frac{1}{3}D_2 \left[ \frac{1}{9} D_1 Z_1.+x.Z_1
\right]\right)$$ \b\;\;\;\;\;\;\left( \bar Z_{2}-\frac{1}{2}
D_{1}\left[Z_2.\bar
\partial+2x.Z_2\right]\right),\e
and $\phi_2$ is a massless minimally coupled scalar field. We should
briefly recall Gupta-Bleuler quantization of the massless minimally
coupled scalar field \cite{gareta}
$$\Box_H \phi(X)=0.$$ As proved by Allen \cite{al}, the
covariant canonical quantization procedure with positive norm states
fails in this case. The Allen's result can be reformulated in the
following way: the Hilbert space generated by a complete set of
modes (named here the positive modes, including the zero mode) is
not dS-invariant, $${\cal H}=\{\sum_{k \geq 0}\alpha_k\phi_k;\;
\sum_{k \geq 0}|\alpha_k|^2<\infty\},$$ where $\phi_k$ is defined in
\cite{gareta}. This means that it is not closed under the action of
the de~Sitter group. Nevertheless, one can obtain a fully covariant
quantum field by adopting a new construction \cite{gareta,dere}. In
order to obtain a fully covariant quantum field, we add all the
conjugate modes to the previous ones. Consequently, we have to deal
with an orthogonal sum of a positive and negative inner product
space, which is closed under an indecomposable representation of the
de~Sitter group. The negative values of the inner product are
precisely produced by the conjugate modes:
$\le\phi_k^*,\phi_k^*\re=-1$, $k\geq 0$. We do insist on the fact
that the space of solution should contain the unphysical states with
negative norm. Now, the decomposition of the field operator into
positive and negative norm parts reads \b
\phi(X)=\frac{1}{\sqrt{2}}\left[ \phi_p(X)+\phi_n(X)\right],\e where
\b \phi_p(X)=\sum_{k\geq 0}
a_{k}\phi_{k}(X)+H.C.,\;\;\phi_n(X)=\sum_{k \geq 0}
b_{k}\phi_k^*(x)+H.C..\e The positive mode $\phi_p(X)$ is the scalar
field as was used by Allen. The crucial departure from the standard
QFT based on CCR lies in the following requirement on commutation
relations: \b    a_{k}|\Omega>=0,\;\;[a_{k},a^{\dag} _{k'}]=
\delta_{kk'},\;\; b_{k}|\Omega>=0,\;\;[b_{k},b^{\dag} _{k'}]=
-\delta_{kk'},\e where $\mid \Omega  \rangle$ is the Gupta-Bleuler
vacuum state. In the next section the Gupta-Bleuler vacuum state is
used to calculated the two-point function of the physical part of
linear gravity.

 \setcounter{equation}{0}
\section{Two-point function}

In the course of intensive studies by various scientists the
following  modalities related to linear gravity have been suggested.
In the main stream approach, it has been found the graviton
propagator in the linear approximation for largely separated points
has a pathological behavior (infrared divergence) and violates the
dS invariance \cite{altu, flilto,anmo1}. Some authors have suggested
that infrared divergence could be exploited in order to create the
instability of the dS space \cite{for, anilto1}. Tsamis and Woodard
have considered a field operator for linear gravity in dS space in
terms of flat coordinates \cite{tswo}, which this coordinates cover
only one-half of the dS hyperboloid. They have examined the
possibility of quantum instability and they have found a quantum
field, which breaks dS invariance.

Antoniadis, Iliopoulos and Tomaras \cite{anilto2} have shown that
the pathological large-distance behavior of the graviton propagator
on a dS background does not manifest itself in the quadratic part of
the effective action  in the one-loop approximation. This means that
the pathological behavior of the graviton propagator may be gauge
dependent and so should not appear in an effective way as a physical
quantity. Recently this result has been also confirmed by several
authors \cite{hiko1, hiko2,ta, vera, ta2, hahetu}.

The important result of the method used in this paper, {\it i.e.}
using the Gupta-Bleuler vacuum, is the calculation of the physical
graviton two-point function, that is dS-invariant and free of any
divergences. In appendix F, the graviton two-point function is
expressed in terms of the dS intrinsic coordinates, which is also
dS-invariant and free of any divergences. This two-point function
can be used for calculation of quantum effects of gravity in the
interaction cases.

Pursuing our method, the two-point function ${\cal W}$, is defined
by \cite{gagata} \b {\cal W}_{\alpha\beta
\alpha'\beta'}(x,x')=\langle
\Omega|\K_{\alpha\beta}(x)\K_{\alpha'\beta'}(x')|\Omega  \rangle ,\e
where $x,x'\in X_H$. This function which is a solution of the wave
Eq.$(3.19)$ with respect to $x$ or $x'$, can be found simply in
terms of the scalar two-point function. We consider the following
possibility for the transverse two-point function
 \b {\cal W}(x,x')=\theta \theta'{\cal W}_0(x,x')+{\cal
S}{\cal S}'\theta.\theta'{\cal W}_{1}(x,x')+D_2D'_2{\cal
W}_g(x,x'),\e where $D_2D'_2=D'_2D_2$ and ${\cal W}_{1}$ and ${\cal
W}_{g}$ are transverse functions. At this stage it is shown that
calculation of ${\cal W}(x,x')$ could be initiated from either $x$
or $x'$ without any difference that means each choices result in the
same equation for ${\cal W}(x,x')$. We first consider the choice
$x$. In this case ${\cal W}(x,x')$ must satisfy the Eq.$(3.19)$,
therefor it is easy to show that: \b \left\{
\begin{array}{ll}
          (Q_0+6)\theta'{\cal W}_0=-4{\cal S}'\theta'.{\cal W}_{1},\;\;\;\;\;\;(I)\\
          \\(Q_1+2){\cal W}_{1}=0, \;\;\;\;\;\;\;\;\;\;\;\;\;\;\;\;\;\;\;\;\;\;\;\;(II)\\
          \\(Q_1+6)D'_2{\cal W}_g=2{\cal S}'(x.\theta'){\cal W}_{1}.\;\;\;\;\;(III)
        \end{array}\right.\e  Using the condition $\partial.{\cal W}_{1}=0,$  Eq.$(5.3-I)$
leads to \b \theta'{\cal W}_0(x,x')=-\frac{2}{3}{\cal
S}'\theta'.{\cal W}_{1}(x,x').\e The solution of the Eq.$(5.3-II)$
can be written as the combination of two arbitrary scalar two-point
functions ${\cal W}_{2}$ and ${\cal W}_{3}$ in the following form
$${\cal W}_{1}=\theta.\theta'{\cal W}_{2}+D_1D'_1{\cal W}_{3}.$$
Substituting this in Eq.$(4.3-II)$ and using the divergenceless
condition we have
$$D'_1{\cal W}_{3}=-\frac{1}{2}\left[2(x.\theta'){\cal W}_{2}+\theta'.\bar{\partial}{\cal W}_{2}\right],$$
$$Q_0{\cal W}_{2}=0.$$
This means that ${\cal W}_{2}$ is the massless minimally coupled
two-point function. Putting ${\cal W}_{2}\equiv{\cal W}_{mc},$ we
have \b {\cal W}_{1}(x,x')=\left(\theta.\theta'
    -\frac{1}{2}D_{1}[\theta'.\bar \partial+2
x.\theta']\right){\cal W}_{mc}(x,x').\e Similar to (4.11) using the
following identity
$$(Q_0+6)^{-1}(x.\theta'){\cal W}_{1}=\frac{1}{6}\left[(x.\theta'){\cal W}_{1}+\frac{1}{9}D_1(\theta'.{\cal W}_{1})\right],$$ the Eq.$(5.3-III)$
leads to
 \b D'_2{\cal W}_g(x,x')=\frac{1}{3} {\cal
S'}\left(\frac{1}{9}D_1\theta'.+x.\theta'\right){\cal
W}_{1}(x,x').\e

According to Eq.s $(5.4)$, $(5.5)$ and $(5.6)$ it turns out that the
two-point function can be written in the following form \b {\cal
W}_{\alpha\beta \alpha'\beta'}(x,x')=\Delta_{\alpha\beta
\alpha'\beta'} (x,\partial,x',\partial'){\cal W}_{mc}(x,x'), \e
where
$$ \Delta_{\alpha\beta \alpha'\beta'}(x,\partial,x',\partial')=-\frac{2}{3}{\cal S'}\theta
\theta'.\left(\theta.\theta'-\frac{1}{2} D_1[\theta'.\bar
\partial+2 \theta'.x]\right)$$
$$ +{\cal S}{\cal S}'\theta.\theta'\left(\theta.\theta'
    -\frac{1}{2}D_1[\theta'.\bar \partial+2
\theta'.x]\right)$$ \b +\frac{1}{3} D_2{\cal
S}'\left(\frac{1}{9}D_1\theta'.+x.\theta'\right)\left(\theta.\theta'
    -\frac{1}{2}D_1[\theta'.\bar \partial+2
\theta'.x]\right).\e

On the other hand with the choice $x'$, the two-point function
$(5.2)$ satisfies  Eq.$(3.19)$ (with respect to $x'$), and we
obtain:
$$ \left\{
\begin{array}{ll}
          (Q'_0+6)\theta{\cal W}_0=-4{\cal S}\theta.{\cal W}_{1},\;\;\;\;\;\;(I)\\
          \\(Q'_1+2){\cal W}_{1}=0, \;\;\;\;\;\;\;\;\;\;\;\;\;\;\;\;\;\;\;\;\;\;(II)\\
          \\(Q'_1+6)D_2{\cal W}_g=2{\cal S}(x'.\theta){\cal W}_{1}.\;\;\;\;\;(III)
        \end{array}\right.$$
Using the condition $\partial'.{\cal W}_{1}=0,$ we have
$$ \theta{\cal W}_0(x,x')=-\frac{2}{3}{\cal S}\theta.{\cal
W}_{1}(x,x') ,$$ $$ D_2{\cal W}_g(x,x')=\frac{1}{3} {\cal
S}\left(\frac{1}{9}D'_1\theta.+x'.\theta\right){\cal W}_{1}(x,x'),$$
$$ {\cal W}_{1}(x,x')=\left(\theta'.\theta
    -\frac{1}{2}D'_{1}[\theta.\bar {\partial'}+2
x'.\theta]\right){\cal W}_{mc}(x,x'),$$ where the primed operators
act on the primed coordinates only. In this case, the two-point
function can be written in the following form $$ {\cal
W}_{\alpha\beta \alpha'\beta'}(x,x')=\Delta'_{\alpha\beta
\alpha'\beta'} (x,\partial,x',\partial'){\cal W}_{mc}(x,x'),
$$ where
$$ \Delta'_{\alpha\beta \alpha'\beta'}(x,\partial,x',\partial')=-\frac{2}{3}{\cal S}\theta
\theta'.\left(\theta.\theta'-\frac{1}{2} D'_1[\theta.\bar
\partial'+2 \theta.x']\right)$$
$$ +{\cal S}{\cal S}'\theta.\theta'\left(\theta.\theta'
    -\frac{1}{2}D'_1[\theta.\bar \partial'+2
\theta.x']\right)$$ $$ +\frac{1}{3} D'_2{\cal
S}\left(\frac{1}{9}D'_1\theta.+x'.\theta\right)\left(\theta.\theta'
    -\frac{1}{2}D'_1[\theta.\bar \partial'+2
\theta.x']\right).$$ In a few steps ahead, it is shown that this
equation is non other than  Eq.$(5.8)$.

The minimally coupled scalar field two-point function in the
Gupta-Bleuler vacuum is \cite{ta3}:
 \b {\cal W}_{mc}(x,x')=\frac{i}{8\pi^2} \epsilon
(x^0-x'^0)[\delta(1-{\cal Z}(x,x'))+\vartheta ({\cal Z}(x,x')-1)],
\e with \b \epsilon (x^0-x'^0)=\left\{ \ba{rcl} 1&x^0>x'^0 ,\\
0&x^0=x'^0 ,\\ -1&x^0<x'^0.\\ \ea\right.\e

Eq.s $(5.4)$, $(5.5)$, $(5.6)$ and $(5.9)$ after relatively simple
and straightforward calculations can be written as (Appendix A):
\b\theta'_{\alpha'\beta'}{\cal W}_0(x,x') =\frac{1}{3}{\cal
S}'\left[\theta'_{\alpha'\beta'}
+\frac{4}{1-{\cal{Z}}^2}(x.\theta'_{\alpha'})(x.\theta'_{\beta'})\right]{\cal{Z}}\frac{d
}{d{\cal{Z}}}{\cal W}_{mc}({\cal{Z}}),\e \b{\cal W}_{1\beta
\beta'}(x,x')=\frac{1}{2}\left[\frac{3+{\cal{Z}}^2}{1-{\cal{Z}}^2}(x'.\theta_{\beta})(x.\theta'_{\beta'})
-{\cal{Z}}(\theta_{\beta}.\theta'_{\beta'})\right]\frac{d
}{d{\cal{Z}}}{\cal W}_{mc}({\cal{Z}}),\e
$$D_{2\alpha}D'_{2\alpha'}{\cal
W}_{g\beta\beta'}(x,x')=-\frac{1}{54(1-{\cal{Z}}^2)^2}{\cal S}{\cal
S}'
[{\cal{Z}}(1-{\cal{Z}}^2)(1+3{\cal{Z}}^2)\theta_{\alpha\beta}\theta'_{\alpha'\beta'}$$
$$+\frac{12{\cal{Z}}}{1-{\cal{Z}}^2}(21-2{\cal{Z}}^2-3{\cal{Z}}^4)(x'.\theta_{\alpha})
(x'.\theta_{\beta})(x.\theta'_{\alpha'})(x.\theta'_{\beta'})
$$
$$+12{\cal{Z}}(1+{\cal{Z}}^2)\theta'_{\alpha'\beta'}(x'.\theta_{\alpha})
(x'.\theta_{\beta})+24{\cal{Z}}(2-{\cal{Z}}^2)\theta_{\alpha\beta}(x.\theta'_{\alpha'})(x.\theta'_{\beta'})$$
$$+{\cal{Z}}(1-{\cal{Z}}^2)(17-9{\cal{Z}}^2)(\theta_{\alpha}.\theta'_{\alpha'})(\theta_{\beta}.\theta'_{\beta'})$$
\b-(79+62{\cal{Z}}^2-45{\cal{Z}}^4)(\theta_{\alpha}.\theta'_{\alpha'})(x.\theta'_{\beta'})(x'.\theta_{\beta})
]\frac{d }{d{\cal{Z}}}{\cal W}_{mc}({\cal{Z}}),\e where $$ Q_0{\cal
W}_{mc}({\cal{Z}})=\frac{3i}{8\pi^2} \epsilon
(x^0-x'^0)[(1-{\cal{Z}})\delta({1-\cal{Z}})]=0. $$

Substituting Eq.s$(5.11)$, $(5.12)$ and $(5.13)$ in $(5.2)$ yields
$$ {\cal W}_{\alpha\beta
\alpha'\beta'}(x,x')=-\frac{2{\cal{Z}}}{27(1-{\cal{Z}}^2)^2}{\cal
S}{\cal
S}'\left[(1-{\cal{Z}}^2)(3{\cal{Z}}^2-2)\theta_{\alpha\beta}\theta'_{\alpha'\beta'}\right.$$
$$+3(1+{\cal{Z}}^2)\theta'_{\alpha'\beta'}(x'.\theta_{\alpha})
(x'.\theta_{\beta})+3(1+{\cal{Z}}^2)\theta_{\alpha\beta}(x.\theta'_{\alpha'})(x.\theta'_{\beta'})$$
$$+\frac{3}{1-{\cal{Z}}^2}(21-2{\cal{Z}}^2-3{\cal{Z}}^4)(x'.\theta_{\alpha})
(x'.\theta_{\beta})(x.\theta'_{\alpha'})(x.\theta'_{\beta'})
$$$$+(1-{\cal{Z}}^2)(11-9{\cal{Z}}^2)(\theta_{\alpha}.\theta'_{\alpha'})(\theta_{\beta}.\theta'_{\beta'})$$
\b\left.-\frac{2}{{\cal{Z}}}(20+{\cal{Z}}^2-9{\cal{Z}}^4)(\theta_{\alpha}.\theta'_{\alpha'})(x.\theta'_{\beta'})(x'.\theta_{\beta})\right]\frac{d
}{d{\cal{Z}}}{\cal W}_{mc}({\cal{Z}}),\e in which we have \b \frac{d
}{d{\cal{Z}}}{\cal W}_{mc}({\cal{Z}})=\frac{i}{8\pi^2}
\frac{{\cal{Z}}-2}{{\cal{Z}}-1}\epsilon
(x^0-x'^0)\delta({\cal{Z}}-1).\e
 Eq. $(5.14)$ is the
explicit form of the two-point function in ambient space notations.
This equation satisfies the traceless and divergenceless conditions:
$$\bar{\partial}.{\cal
W}=\bar{\partial'}.{\cal W}=0,\;\;\;\mbox{and}\;\;\;{\cal
W}_{\alpha\beta \alpha'}^{\;\;\;\;\alpha'}(x,x')={\cal
W}^{\alpha}_{\alpha \alpha'\beta'}(x,x')=0.$$   The two-point
function $(5.14)$ is obviously dS-invariant and free of any
divergences. The ambient space notation clearly exhibits this fact
that the gravitational field, ${\cal K}_{\alpha\beta}$, can be
written in terms of the minimally coupled scalar field directly
eq. (4.13). It should be noted that the Gupta-Bleuler quantization
of minimally coupled scalar field, irrespective of choice of
ambient space notation, does completely eliminate the infrared
divergence in the scalar two-point function \cite{gareta}. In
Appendix F, the intrinsic counterpart of (5.14) is calculated.

\section{Conclusion}

It was pointed out that Einstein's theory of gravitation, in the
background field method, $g_{\mu\nu}=g_{\mu\nu}^{BG}+h_{\mu\nu}$,
can be considered as a massless symmetric tensor field of rank-2 on
a fixed background, such as dS space. Contrary to  Maxwell equation,
the Einstein's equation of gravitation, as well as equation of
$h_{\mu\nu}$, is not conformally invariant.

In this paper we used Dirac's six-cone formalism to obtain CI
"massless" spin-2 wave equation in dS space which corresponds to
UIR's of dS group ($n=2$ in (3.3)). It was shown that the intrinsic
counterpart of CI wave equation with conformal degree $-1$ ($n=1$ in
(3.3)) is similar to what Barut and Xu have obtained in Minkowski
space. This, however, is not a physical state of dS group. Barut and
Bohm \cite{ms9} have shown that for the physical representation of
the conformal group, the value of the conformal Casimir operator is
$9$. But according to (3.1) for the tensor field of rank-2 and
conformal degree 0, this value becomes 8 on the cone. Therefore
tensor field of rank-2 does not correspond to the UIR's of the
conformal group (physical state of group). In other words, the
tensor field that carries the physical representations of conformal
group must be a tensor field of higher rank. In the forthcoming
paper we will consider a mixed symmetry rank-3 tensor field
$\Psi^{abc}$ with degree zero that transforms simultaneously under
the action of dS and conformal groups.

In addition to obtain the CI wave equation in dS space, we have
shown that the physical part of the linear gravity, in ambient space
notations, can be written as the product of a generalized symmetric
rank-2 polarization tensor and a massless minimally coupled scalar
field. Using the Gupta-Bleuler quantization we have calculated the
physical graviton two-point function, which is dS-invariant and free
of any divergences. This two-point function can be used to calculate
the quantum effects of gravity in the interaction cases, which will
be considered in forthcoming papers.

\vspace{0.5cm} \noindent {\bf{Acknowlegements}}: The authors would
like to thank S. Fatemi for her interest in this work.

\setcounter{equation}{0}
\begin{appendix}
\section{Some useful relations}

In this appendix, some useful relations are given. The action of the
Casimir operators $Q_1 $ and $ Q_2$ can be written in the more
explicit form \b Q_1
K_{\alpha}=(Q_0-2)K_{\alpha}+2x_{\alpha}\partial . K
-2\partial_{\alpha} x\cdot K, \e \b Q_2 {\cal K}_{\alpha
\beta}=(Q_0-6){\cal K}_{\alpha \beta}+2{\cal
S}x_{\alpha}\partial\cdot {\cal K}_{\beta}-2{\cal
S}\partial_{\alpha}x\cdot {\cal
K}_{\beta}+2\eta_{\alpha\beta}{\K}'\, \e \b Q_1 D_1 =D_1 Q_0 ,\e \b
(Q_{0}-2)x_{\alpha}=x_{\alpha}Q_{0}-6x_{\alpha}-2\bar\partial_{\alpha},
\e \b
\bar\partial_{\alpha}(Q_{0}-2)=Q_{0}\bar\partial_{\alpha}-8\bar\partial_{\alpha}-
2Q_{0}x_{\alpha} - 8x_{\alpha},\e \b
x_{\alpha}Q_{0}(Q_{0}-2)=(Q_{0}-2)(Q_{0}x_{\alpha}+4x_{\alpha}+4\bar\partial
_{\alpha}), \e  \b [Q_{0}Q_{2},Q_{2}Q_{0}]{\cal
K}_{\alpha\beta}=4{\cal
S}(x_{\alpha}-\bar\partial_{\alpha})\bar\partial .{\cal
K}_{\beta},\e   the transverse divergence $\bar
\partial _{\alpha}$ can be written with respect to
$\partial_{\alpha}$ as the following \b \bar \partial _{\alpha}
\equiv \partial_{\alpha}+x_{\alpha}x\cdot
\partial=\partial_{\alpha}-x_{\alpha}+x\cdot \partial
x_{\alpha}. \e

To obtain the two-point function, the following identities are used
 \b \bar{\partial}_\alpha
f({\cal{Z}})=-(x'.\theta_{\alpha})\frac{d f(\cal{Z})}{d{\cal{Z}}},\e
\b\theta^{\alpha\beta}\theta'_{\alpha\beta}=\theta..\theta'=3+{\cal{Z}}^2,
\e \b (x.\theta'_{\alpha'})(x.\theta'^{\alpha'})={\cal{Z}}^2-1,\e \b
(x.\theta'_{\alpha})(x'.\theta^{\alpha})={\cal{Z}}(1-{\cal{Z}}^2),\e
\b\bar{\partial}_\alpha(x.\theta'_{\beta'})=\theta_{\alpha}.\theta'_{\beta'},\e
\b\bar{\partial}_\alpha(x'.\theta_{\beta})=x_\beta(x'.\theta_{\alpha})-{\cal{Z}}\theta_{\alpha\beta},\e
\b\bar{\partial}_\alpha(\theta_{\beta}.\theta'_{\beta'})=x_\beta(\theta_{\alpha}.\theta'_{\beta'})+
\theta_{\alpha\beta}(x.\theta'_{\beta'}),\e
\b\theta'^{\beta}_{\alpha'}(x'.\theta_{\beta})=-{\cal{Z}}(x.\theta'_{\alpha'}),\e
\b\theta'^{\gamma}_{\alpha'}(\theta_{\gamma}.\theta'_{\beta'})=\theta'_{\alpha'\beta'}+(x.\theta'_{\alpha'})(x.\theta'_{\beta'}),\e
\b Q_0f({\cal{Z}})=(1-{\cal{Z}}^2)\frac{d^2
f(\cal{Z})}{d{\cal{Z}}^2}-4{\cal{Z}}\frac{d
f(\cal{Z})}{d{\cal{Z}}}.\e

\setcounter{equation}{0}
\section{CI wave equation with $n=1$}

We show that the CI wave equation for the tensor field $\Psi_{ab}$
with $n=1$, doesn't transform according to the UIR's of the dS and
conformal groups.

The CI system $(3.3)$ with $n=1$, {\it i.e.} for the tensor field
with degree $-1$, reads as \b (Q_0-2)\Psi_{\alpha\beta}=0, \e using
the transversality condition, $u_a \Psi^{ab}=0,$ we get \b
(Q_0-2)x.\Psi_{\beta}=0, \e \b (Q_0-2)x.\Psi.x=0.\e note that the
relation (3.3) is used.

Multiplying (B.1)and (B.2) by $x_{\beta}$ result in \b
2x.\Psi_{\alpha}=-\bar\partial\cdot \Psi_{\alpha}, \e \b
2x.\Psi.x=-\bar\partial \cdot \Psi.x\,. \e The divergence of
$\K_{\alpha\beta}$ leads to \b
\bar\partial\cdot\K_{\beta}=\bar\partial\cdot\Psi_{\beta}+5x.\Psi_{\beta}+5x_{\beta}x.\Psi.x+
x_{\beta}\bar\partial\cdot\Psi\cdot x\,. \e Combining the Eq.(B.6)
and (B.5),(B.4), leads to  \b \bar\partial \cdot
\K_{\beta}=3(x.\Psi_{\beta}+x_{\beta}x.\Psi.x)\,. \e

After some calculations one finds $$
(Q_0-2)\K_{\alpha\beta}+2(\bar\partial
_{\alpha}+2x_{\alpha})x.\Psi_{\beta}+2(\bar\partial_{\beta}+2x_{\beta})x.\Psi_{\alpha}+2(\bar\partial
_{\alpha}+2x_{\alpha})x_{\beta}x.\Psi.x$$ \b  +
2x_{\alpha}(\bar\partial_{\beta}+2x_{\beta})x\cdot\Psi\cdot x=0\,.\e
Substituting Eq.(B.7) into Eq.(B.8), leads exactly to (3.10). In
order to express Eq.(3.9) in terms of the intrinsic coordinates the
following relation is used:  \cite{ta}$$ \nabla_\mu \nabla_\nu
\cdots \nabla_\rho h_{\lambda_{1} \cdots \lambda_{l}}=\frac{\partial
x^\alpha}{\partial X^\mu}\frac{\partial x^\beta}{\partial X^\nu}
\cdots \frac{\partial x^\gamma}{\partial X^\rho}\frac{\partial
x^{\eta_1}}{\partial X^{\lambda_1}}\cdots \frac{\partial
x^{\eta_l}}{\partial X^{\lambda_l}} Trpr \bar\partial_\alpha
Trpr\bar\partial_\beta \cdots Trpr \bar\partial_\gamma \K_{\eta_1
\cdots \eta_l}$$ where the transverse projection defined by $$ (Trpr
\K)_{\lambda_1 \cdots \lambda_l}\equiv \theta_{\lambda_1}^{\eta_1}
\cdots\theta_{\lambda_l}^{\eta_l}\K_{\eta_{1 \cdots \eta_l}}
$$ guarantees the transversality in each index. Applying this
procedure to a transverse second rank, symmetric tensor field, leads
to $$ \nabla_\mu \nabla_\nu h_{\rho \lambda}=\frac{\partial
x^\alpha}{\partial X^\mu}\frac{\partial x^\beta}{\partial
X^\nu}\frac{\partial x^\gamma}{\partial X^\rho}\frac{\partial
x^{\eta}}{\partial X^\lambda} Trpr \bar\partial_\alpha Trpr
\bar\partial_\beta \K_{\gamma\eta}$$ where we have
$$
Trpr \bar\partial_\alpha Trpr \bar\partial_\beta
\K_{\gamma\eta}=\bar\partial_\alpha(\bar\partial_\beta
\K_{\gamma\eta}-x_\gamma \K_{\beta\eta}-x_\eta
\K_{\gamma\beta})$$$$-x_\beta (\bar\partial_\alpha
\K_{\gamma\eta}-x_\gamma \K_{\alpha \eta}-x_\eta
\K_{\gamma\alpha})-x_\gamma (\bar\partial_\beta
\K_{\alpha\eta}-x_\alpha \K_{\beta \eta}-x_\eta
\K_{\alpha\beta})-x_\eta (\bar\partial_\beta-x_\gamma
\K_{\beta\alpha}-x_\alpha \K_{\gamma\beta}).$$

Thus we can write\b \nabla_\lambda\nabla^\lambda
h_{\mu\nu}\equiv\Box h_{\mu\nu}\rightarrow \bar\partial_\alpha
\bar\partial^\alpha \K_{\gamma\eta}-2\K_{\gamma\eta}-2{\cal S}
x_\gamma \bar\partial\cdot \K_\eta\,,\e \b \nabla_\lambda \nabla
\cdot h_\mu\rightarrow \bar\partial_\eta \bar\partial \cdot
\K_\gamma-x_\gamma\bar\partial\cdot \K_\eta\,, \e
$$g^{dS}_{\mu\nu}\rightarrow \theta_{\gamma\eta}$$

Using the above statements and $Q_0=-\bar\partial_\alpha
\bar\partial^\alpha$ the intrinsic counterpart of (3.10) can be
easily derived.

\setcounter{equation}{0}
\section{Some details about Eq.s (3.15) and (3.16)}

The condition $(3.14)$ for the tensor field with degree zero leads
to \b \partial . \Psi_{\beta}=-x \cdot\partial
x\cdot\Psi_{\beta}\,,\e \b \partial . \Psi_5 =-x\cdot\partial
x\cdot\Psi_5\,.\e Combining (C.2) and $(3.13)$ results in \b
\partial .\Psi . x+x\cdot \partial x.\Psi .x=0\,.\e
In this case we rewrite (B.6) in the following form \b \bar
\partial .
\K_{\beta}=4(\Psi_{\beta}.x+x_{\beta}x.\Psi.x)+(\bar\partial
.\Psi_{\beta}+x. \Psi_{\beta}+x_{\beta}x.\Psi.x
+x_{\beta}\bar\partial.\Psi.x)\,.\e According to relations
(A.8),(C.1) and (C.3), the second parenthesis vanishes, therefore we
get Eq.(3.15).

Finally according to (A.1) and (3.15), we can write the following
relations for the vector field $\bar\partial . \K_{\alpha}$ \b
Q_1\bar\partial.\K_{\alpha}=(Q_0-2)\bar\partial.
\K_{\alpha}+2x_{\alpha}\bar\partial .\bar\partial . \K\,,\e \b
(Q_0-2)\bar\partial
.\K_{\alpha}=4((Q_{0}-2)x.\Psi_{\alpha}+(Q_0-2)x_{\alpha}x.\Psi.x)\,.\e
After some calculation it is easy to show that \b
(Q_0-2)(x.\Psi_{\alpha}+x_{\alpha}x.\Psi.x)=-\frac{1}{6}(D_{1\alpha}Q_0+4D_{1\alpha}+12x_{\alpha})
(\bar\partial.\Psi.x+4x.\Psi.x)\,.\e Note that \b \bar\partial.
\bar\partial.\K=4(\bar\partial.\Psi.x+4x.\Psi.x)\,. \e By
substituting (C.8) and (3.15) into (C.7), we get (3.16).

\setcounter{equation}{0}
\section{Details of calculation of Eq.(3.17)}

For symmetric rank-2 field $\Psi_{ab}$, the CI system (3.12) results
in
\b \left.\ba{rcl}  Q_{0}(Q_{0} -2)\Psi _{\alpha\beta}&=&0,\\
Q_{0}(Q_{0} -2)\Psi_{55}&=&0. \ea\right. \e Using conditions $\K'=0$
and (3.13), we get \b Q_0(Q_0-2)x.\Psi.x=0\,,\e \b
Q_0(Q_0-2)x.\Psi_{\beta}=0\,.\e Taking the divergence of (3.16)
leads to \b Q_0(Q_0-2)\bar\partial.\Psi.x=0\,.\e The action of
operator $Q_0(Q_0-2)$ on dS field can be written in more explicit
form \b Q_0(Q_0-2)\K_{\alpha\beta}=Q_{0}(Q_{0}-2) {\cal S}
x_{\alpha} \Psi_{\beta} \cdot x + Q_{0}(Q_{0}-2)x_{\alpha}x_{\beta}x
\cdot \Psi \cdot x\, .\e According to (A.4) and (A.5), the above
equation can be written as follows
$$Q_{0}(Q_{0}-2){\cal
K}_{\alpha\beta}=-4(3x_{\alpha}+\bar\partial_{\alpha})(Q_{0}-2)\Psi_{\beta}
\cdot x $$ \b -4
(3x_{\beta}+\bar\partial_{\beta})(Q_{0}-2)\Psi_{\alpha}\cdot x
-4x_{\alpha}(3x_{\beta}+\bar\partial_{\beta})(Q_{0}-2)x \cdot \Psi
\cdot x -4 (3x_{\alpha}+\bar\partial_{\alpha})(Q_{0}-2)x_{\beta}x
\cdot \Psi \cdot x, \e or we can write $$ Q_{0}(Q_{0}-2){\cal
K}_{\alpha\beta}=-4(3x_{\beta}+\bar\partial_{\beta})(Q_{0}-2)\Psi_{\alpha}\cdot
x  $$ \b -(3x_{\alpha}+\bar\partial_{\alpha})(Q_{0}-2)\bar\partial
\cdot {\cal
K}_{\beta}-4x_{\alpha}(3x_{\beta}+\bar\partial_{\beta})(Q_{0}-2)x
\cdot \Psi \cdot x\,\,,\e note that identity (3.15) is used.

Multiplying (D.3) by $x_{\beta}$ results in \b
(Q_0-2)(x.\Psi.x+\bar\partial.\Psi.x)=0\,.\e Substituting the
divergence of (3.15) into the above equation leads to \b
(Q_0-2)x.\Psi.x=\frac{1}{12}(Q_0-2)\bar\partial\cdot
\bar\partial\cdot\K\,.\e  So we can rewrite(D.7) as follows
$$(Q_{0}-2)Q_{0}{\cal
K}_{\alpha\beta}=-4(3x_{\beta}+\bar\partial_{\beta})(Q_{0}-2)\Psi_{\alpha}\cdot
x $$ \b -(3x_{\alpha}+\bar\partial_{\alpha})(Q_{0}-2)\bar\partial
\cdot {\cal
K}_{\beta}-\frac{1}{3}x_{\alpha}(3x_{\beta}+\bar\partial_{\beta})(Q_{0}-2)\bar\partial
. \bar \partial .{\cal K}\,\,.\e Multiplying the above equation by
$x_{\beta}$ leads to \b (Q_{0}-2)\Psi_{\alpha}\cdot
x=\frac{1}{4}(Q_{0}-2)\bar\partial\cdot {\cal K}_{\alpha}
+\frac{1}{3}x_{\alpha}\bar\partial \cdot \bar\partial\cdot {\cal
K}-\frac{1}{12}x_{\alpha}(Q_{0}-2)\bar\partial \cdot
\bar\partial\cdot {\cal
K}+\frac{1}{6}\bar\partial_{\alpha}\bar\partial \cdot
\bar\partial\cdot {\cal K}\,\,.\e Finally combining (D.11) and
(D.10) leads to $$ (Q_{0}-2)Q_{0}{\cal K}_{\alpha\beta} +Q_{0}{\cal
S}x_{\beta}\bar\partial \cdot {\cal K}_{\alpha}+Q_{0}{\cal
S}\bar\partial_{\alpha}\bar\partial \cdot {\cal K}_{\beta}-2{\cal
S}x_{\alpha}\bar\partial \cdot {\cal K}_{\beta}-2{\cal
S}\bar\partial_{\alpha}\bar\partial \cdot {\cal K}_{\beta}$$ $$
+4x_{\alpha}x_{\beta}\bar\partial . \bar\partial . {\cal K}
+\frac{1}{3} {\cal
S}\bar\partial_{\beta}\bar\partial_{\alpha}\bar\partial .
\bar\partial . {\cal K}+\frac{5}{3}{\cal
S}x_{\alpha}\bar\partial_{\beta}\bar\partial . \bar\partial . {\cal
K}$$ \b +2\theta_{\alpha\beta}\bar\partial . \bar\partial . {\cal
K}-\frac{1}{3}\theta_{\alpha\beta}Q_{0}\bar\partial . \bar\partial .
{\cal K}=0\,\,,\e

It is easy to show that if we rewrite Eq.(3.17) in terms of $Q_0$,
we will get back exactly to  Eq.(D.12). Note that for this
calculation the following relations have been used
$$(Q_{2}+4)(Q_{2}+6){\cal K}_{\alpha\beta}=$$ \b
Q_{0}(Q_{0}-2){\cal K}_{\alpha\beta}+4{\cal
S}[Q_{0}x_{\alpha}\bar\partial . {\cal
K}_{\beta}+3x_{\alpha}\bar\partial . {\cal
K}_{\beta}+\bar\partial_{\alpha} \bar\partial . {\cal
K}_{\beta}+x_{\alpha}x_{\beta}\bar\partial . \bar\partial . {\cal
K}] \,\,, \e

$$ (Q_{2}+4)D_{2\alpha}\partial_2 . {\cal
K}_{\beta}={\cal S} [-3Q_{0}x_{\alpha} \bar\partial . {\cal
K}_{\beta}+Q_{0} \bar\partial_{\alpha} \bar\partial . {\cal
K}_{\beta} $$ \b -6 \bar\partial_{\alpha} \bar\partial . {\cal
K}_{\beta}-14x_{\alpha} \bar\partial . {\cal
K}_{\beta}-2x_{\alpha}x_{\beta} \bar\partial . \bar\partial . {\cal
K}+ 2\theta_{\alpha\beta} \bar\partial . \bar\partial . {\cal K}+
2x_{\beta}\bar\partial_{\alpha} \bar\partial . \bar\partial . {\cal
K} ] \,, \e

\b D_{2\alpha}D_{1\beta}\bar\partial . \bar\partial . {\cal K}={\cal
S}[\bar\partial_{\alpha}\bar\partial_{\beta}-x_{\alpha}\bar\partial_{\beta}]\bar\partial
. \bar\partial . {\cal K}\,\,.\e

\setcounter{equation}{0}
\section{Details on equation (4.12)}

Using $(A.1)$, it is easy to show that \b
D_1(Z_1.K)=\frac{1}{6}(Q_1+6)[D_1(Z_1.K)],\e \b
x(Z_1.K)=\frac{1}{6}(Q_1+6)[x(Z_1.K)], \e \b
Z_1.\bar{\partial}K=\frac{1}{6}(Q_1+6)[Z_1.\bar{\partial}K-\frac{1}{3}D_1(Z_1.K)],\e
\b (Q_1+6)[(x.Z_1)K]=2[x(Z_1.K)-Z_1.\bar{\partial}K].\e The
conditions $ x.K=\bar{\partial}.K=0, $ and $Q_0 K=0, $ are used to
obtain the above equations.

Substituting Eq.s $(E.2)$ and $(E.3)$ in $(E.4)$ we have \b
(Q_1+6)[(x.Z_1)K]=\frac{1}{3}(Q_1+6)\left[\frac{1}{3}D_1(Z_1.K)+x(Z_1.K)-Z_1.\bar{\partial}K\right],\e
or \b
(x.Z_1)K=\frac{1}{3}\left[\frac{1}{3}D_1(Z_1.K)+x(Z_1.K)-Z_1.\bar{\partial}K\right];\e

finally according Eq.s $(E.1)$ and $(E.4)$, we obtain
\b(x.Z_1)K=\frac{1}{6}(Q_1+6)\left[\frac{1}{9}D_1(Z_1.K)+(x.Z_1)K
\right].\e This automatically leads to  Eq.$(4.12)$.

 \setcounter{equation}{0}
\section{Two-point function in dS intrinsic coordinates}

In order to compare our results with the work of the other authors
\cite{hiko1,hiko2}, we write the two-point function in dS space
(maximally symmetric) in terms of bitensors. These are functions of
two points $(x,x')$ and behave like tensors under coordinate
transformations at each points.

As mentioned in \cite{gagata}, any maximally symmetric bitensor can
be expressed as a sum of products of three basic tensors. The
coefficients in this expansion are functions of the geodesic
distance $\sigma(x, x')$, that is the distance along the geodesic
connecting the points $x$ and $x'$ (note that $ \sigma(x, x')$ can
be defined by an unique analytic extension also when no geodesic
connects $x$ and $x'$). In this sense, these fundamental tensors
form a complete set. They can be obtained by differentiating the
geodesic distance:
$$n_\mu = \nabla_\mu \sigma(x, x')\;\;\;,\;\;\; n_{\mu'} = \nabla_{\mu'} \sigma(x,
x'), $$and the parallel propagator
$$g_{\mu\nu'}=-c^{-1}({\cal{Z}})\nabla_{\mu}n_{\nu'}+n_\mu n_{\nu'}.$$
The geodesic distance is implicitly defined for $ {\cal{Z}}=-x\cdot
x', $ by: 1) $ {\cal{Z}}=\cosh (\sigma )$ if $x$ and $x'$ are
time-like separated,
 2) $ {\cal{Z}}=\cos (\sigma ) $ if
$x$ and $x'$are space-like separated. The basic bitensors in ambient
space notations are found through
$$ \bar{\partial}_\alpha \sigma(x,x')\;\;\;,\;\;\;\bar{\partial}'_{\beta'}
\sigma(x,x')\;\;\;,\;\;\;\theta_\alpha .\theta'_{\beta'},$$
restricted to the hyperboloid by
$$ {\cal{T}}_{\mu\nu'}=\frac{\partial x^\alpha}{\partial
X^\mu}\frac{\partial x'^{\beta'}}{\partial
X'^{\nu'}}T_{\alpha\beta'}.$$

For $ {\cal{Z}}=\cos(\sigma), $ one can find
$$n_\mu=\frac{\partial x^\alpha}{\partial X^\mu}\bar{\partial}_\alpha \sigma(x,x')=
\frac{\partial x^\alpha}{\partial X^\mu} \frac{(x' \cdot
\theta_\alpha)}{\sqrt{1-{\cal{Z}}^2}},\;\; n_{\nu'}=\frac{\partial
x'^{\beta'}}{\partial X'^{\nu'}}\bar{\partial}_{\beta'} \sigma(x,x')
=\frac{\partial x'^{\beta'}}{\partial X'^{\nu'}}
\frac{(x\cdot\theta'_{\beta'})}{\sqrt{1-{\cal{Z}}^2}},$$
$$\nabla_\mu n_{\nu'}=\frac{\partial x^\alpha}{\partial
X^\mu}\frac{\partial x'^{\beta'}}{\partial
X'^{\nu'}}\theta^\varrho_\alpha
\theta'^{\gamma'}_{\beta'}\bar{\partial}_\varrho\bar{\partial}_{\gamma'}
\sigma(x, x')=c({\cal{Z}})[n_\mu n_{\nu'}{\cal{Z}}-\frac{\partial
x^\alpha}{\partial X^\mu}\frac{\partial x'^{\beta'}}{\partial
X^{\nu'}}\theta_\alpha \cdot\theta'_{\beta'}],$$ with $
c^{-1}({\cal{Z}})=-\frac{1}{\sqrt{1-{\cal{Z}}^2}}.$  For $
{\cal{Z}}=\cosh (\sigma), $  $ n_\mu $ and $ n_\nu$ are multiplied
by $i$ and $ c({\cal{Z}}) $ becomes
$-\frac{i}{\sqrt{1-{\cal{Z}}^2}}.$ In both cases we have
$$ g_{\mu\nu'}+({\cal{Z}}-1)n_\mu n_{\nu'}=\frac{\partial x^\alpha}{\partial
X^\mu}\frac{\partial x'^{\beta'}}{\partial X'^{\nu'}}\theta_\alpha
\cdot\theta'_{\beta'} .$$ and the two-point functions are related
through $$ Q_{\mu\nu\mu'\nu'}= \frac{\partial x^\alpha}{\partial
X^\mu} \frac{\partial x^\beta}{\partial X^\nu} \frac{\partial
x'^{\alpha'}}{\partial X'^{\mu'}} \frac{\partial
x'^{\beta'}}{\partial
X'^{\nu'}}{\cal{W}}_{\alpha\beta\alpha'\beta'}.$$

Considering the above expressions the two-point function (5.14)
takes the following form
$$Q_{\mu\nu\mu'\nu'}(X,X')=-\frac{2}{27(1-{\cal{Z}}^2)}{\cal S}{\cal
S}'\left[{\cal{Z}}(3{\cal{Z}}^2-2)g_{\mu\nu}g'_{\mu'\nu'}+3{\cal{Z}}(1+{\cal{Z}}^2)
(g'_{\mu'\nu'}n_{\mu}n_{\nu}+g_{\mu\nu}n_{\mu'}n_{\nu'})\right.
$$$$+{\cal{Z}}(11-9{\cal{Z}}^2)g_{\mu\mu'}g_{\nu\nu'}+
\left(40+32{\cal{Z}}-20{\cal{Z}}^2-6{\cal{Z}}^3+9{\cal{Z}}^4-9{\cal{Z}}^5
\right) n_{\mu}
n_{\nu}n_{\mu'}n_{\nu'}$$\b+\left.\left(-40+9{\cal{Z}}^2+9{\cal{Z}}^4\right)g_{\mu\mu'}n_\nu
n_{\nu'} \right]\frac{d }{d{\cal{Z}}}{\cal W}_{mc}({\cal{Z}}).\e The
two-point function $(F.1)$ is obviously dS-invariant, and appearance
of the factors $ {\cal{Z}}\; \delta({\cal{Z}}-1)\;,\;{\cal{Z}}^2\;
\delta({\cal{Z}}-1) \;,\;{\cal{Z}}^3\; \delta({\cal{Z}}-1),$ make it
free of any divergences.

\end{appendix}

\end{document}